\documentclass[12pt,a4paper]{article}
\usepackage[T2A]{fontenc}
\usepackage[utf8]{inputenc}
\usepackage[english]{babel}
\usepackage{amssymb}
\usepackage{amsmath}
\usepackage{graphicx}
\usepackage[small,sc,center]{titlesec}
\usepackage{indentfirst}
\usepackage{siunitx}
\usepackage[labelsep=period]{caption}
\usepackage{caption}

\begin{document}

\begin{center}
	\textbf{Theoretical period--radius and period--luminosity relations for Mira variables with solar
	metallicity}

	\vskip 3mm
	\textbf{Yu. A. Fadeyev\footnote{E--mail: fadeyev@inasan.ru}}

	\textit{Institute of Astronomy, Russian Academy of Sciences,
    		Pyatnitskaya ul. 48, Moscow, 119017 Russia} \\

	Received October 16, 2023; revised November 6, 2023; accepted November 6, 2023
\end{center}

\textbf{Abstract} ---
Evolutionary sequences of AGB stars with initial masses on the main sequence
$M_\mathrm{ZAMS}=1.5M_\odot$, $2M_\odot$ and $3M_\odot$ were computed for
the initial metallicity $Z=0.014$.
Selected models of evolutionary sequences with envelopes under thermal equilibrium
were used as initial conditions for calculation of nonlinear stellar pulsations.
The hydrodynamic models of each evolutionary sequence are shown to concentrate
along the continuous line in the period--radius and period--luminosity diagrams.
The theoretical period--radius and period--luminosity relations differ from one another
for different main--sequence star masses
because the stellar luminosity of AGB stars depends on the degenerate carbon core mass
which increases with increasing $M_\mathrm{ZAMS}$.
In hydrodynamic models of evolutionary sequences $M_\mathrm{ZAMS}=2M_\odot$ and
$M_\mathrm{ZAMS}=3M_\odot$ the periods of the first overtone pulsators are
$86~\textrm{d}\le\Pi\le 123~\textrm{d}$ and $174~\textrm{d}\le\Pi\le 204~\textrm{d}$,
whereas all models of the evolutionary sequence $M_\mathrm{ZAMS}=1.5M_\odot$
oscillate in the fundamental mode.
Fairly regular radial oscillations exist in stars with pulsation periods $\Pi\lesssim 500$~d.
In models with longer periods the amplitude rapidly increases with increasing $\Pi$ and
oscillations become irregular.

Keywords: \textit{stellar evolution; stellar pulsation; stars: variable and peculiar}

\newpage
\section*{introduction}

Soon after the discovery of a correlation between the period of light variations and the bolometric
magnitude in eleven Mira pulsating variables of the Large Magellanic Cloud
(Glass and Lloyd Evans 1981) these stars became a reliable distance indicator together with
the classical Cepheids (Glass and Feast 1982а; 1982b; Feast 1985).
The near--infrared radiation at wavelengths $\sim 2~\mu\textrm{m}$ is less sensitive to
the interstellar extinction in comparison with the optical range and due to the large luminosity in
the infrared the distance scale of Miras extends much further in comparison with that of Cepheids.
Moreover, more reliable estimates of bolometric magnitudes provided by the near--infrared photometry
play an important role in comparison of theoretical models with observations of Mira variables.

The period--luminosity relation was first determined from observations of Mira variables in
the Magellanic Clouds (Feast 1984; Feast et al. 1989; Groenewegen and Whitelock 1996).
Somewhat later the similar correlations between the luminosity and the period of light variations
were found for Miras observed in galaxies of the Local Group M31 (Mould et al. 2004),
M33 (Mould et al. 1990; Yuan et al. 2018), NGC~4528 (Huang et al. 2018) and
NGC~6822 (Whitelock et al. 2013).
The period--luminosity relation of galactic Miras is based on measurements of their trigonometric
parallaxes obtained with the VLBI observations (Chibueze et al. 2020; Urago et al. 2020;
Sun et al. 2022) as well as with astrometric  satellites Hipparcos (Whitelock et al. 2000;
Whitelock and Feast 2000) and Gaia (Andriantsaralaza et al. 2022; Zhang and Sanders 2023).

It is intuitively clear that the existence of the period--luminosity relation of Miras is due to
the evolution of AGB stars of various masses, ages and composition.
Therefore, the nature of this relationship can only be understood with the help of consistent
stellar evolution and stellar pulsation computations.
Unfortunately, no detailed consistent computations have been done so far in this field.
The purpose of this work is to fill this gap for the AGB stars with a solar metallicity.
Below we present the results of the Cauchy problem solution for the equations of radiation
hydrodynamics describing the nonlinear stellar pulsations and where the selected AGB stellar
evolution models are used as the initial conditions.
The final goal of hydrodynamic calculations is to determine the mean pulsation period $\Pi$
of the hydrodynamic model after attainment of the limiting amplitude when $\Pi$ is
time--independent.

\section*{evolutionary sequences of agb stars}

In this work we considered the evolution of stars with masses on the main sequence
$M_\mathrm{ZAMS}=1.5M_\odot$, $2M_\odot$ and $3M_\odot$.
The initial abundance of helium was set to be $Y=0.28$ whereas the abundance of heavier elements
was assumed to correspond to the solar metallicity $Z=0.014$ (Asplund et al. 2009).
Calculations of the stellar evolution from the main sequence up to the final AGB stage were done
with the program MESA version r15140 (Paxton et al. 2019).
Convective mixing of the stellar matter was treated via the theory of B\"ohm--Vitense (1958)
for the mixing length to pressure scale height ratio $\alpha_\mathrm{MLT}=1.8$.
Extra mixing due to overshooting at the convection zone boundaries was calculated according to
Herwig (2000) and Pignatari et al. (2016).
The nuclear energy generation rates and the nucleosynthesis were calculated with the help of
the JINA Reaclib database (Cyburt et al. 2010).
The mass loss rate due to the stellar wind was calculated according to Reimers (1975) with
the parameter $\eta_\mathrm{R}=0.5$ at evolutionary stages prior to the AGB
whereas at the AGB phase the mass loss rate was computed using the formula of Bl\"ocker (1995)
with the parameter $\eta_\mathrm{B}=0.05$.
To evaluate the role of uncertainties in the mass loss rates we computed two additional
evolutionary sequences $M_\mathrm{ZAMS}=1.5M_\odot$ and $3M_\odot$ with the mass loss rate
parameter $\eta_\mathrm{B}=0.03$.

The theory of stellar pulsation is only applicable to models of stellar envelopes under both
the hydrostatic and thermal equilibrium.
The first of these conditions is always fulfilled since the solution of the equations of
stellar evolution describes the stellar structure under the hydrostatic equilibrium.
However at some evolutionary stages the stellar envelope is in thermal imbalance
owing to changes of its gravitational energy during envelope contraction or expansion.

The equation of energy conservation for the spherically--symmetric star is
\begin{equation}
\label{enercons1}
\frac{d L_r}{d M_r} = \varepsilon - \varepsilon_\nu - T\frac{\partial S}{\partial t} ,
\end{equation}
where $M_r$ is the Lagrangean coordinate which equals the mass confined to a radius $r$,
$L_r$ is the total (radiative plus convective) luminosity in the layer with radius $r$,
$\varepsilon$ is the nuclear energy generation rate,
$\varepsilon_\nu$ is the cooling rate via neutrino emission,
$T$ and $S$ are the temperature and specific entropy of the stellar matter, $t$ is time.
Analysis of stellar pulsation is usually restricted to outer layers of the red giant with
temperature $T\lesssim 10^7$~K so that the first two terms on the right--hand side of
(\ref{enercons1}) are negligible and can be omitted.
Therefore, the condition of the thermal equilibrium in the spherically--symmetric stellar
envelope is written as
\begin{equation}
\label{enercons2}
\frac{d L_r}{d M_r} = 0 .
\end{equation}
In the present study to measure the deviation from thermal equilibrium we used the parameter
\begin{equation}
\label{deltaL}
\delta_\mathrm{L} = \max_{1\le j\le N} |1 - L_j/L_0| ,
\end{equation}
where $L_j$ is the total luminosity at the $j$--th Lagrangean zone, $j=0$ and $j=N$
correspond to the inner and outer boundaries of the stellar envelope model.
Earlier we have shown (Fadeyev 2022) that the condition $\delta_\mathrm{L}\lesssim 10^{-2}$
is a good approximation of the thermal equilibrium in the stellar model.

\begin{figure}
\centerline{\includegraphics[width=0.9\textwidth]{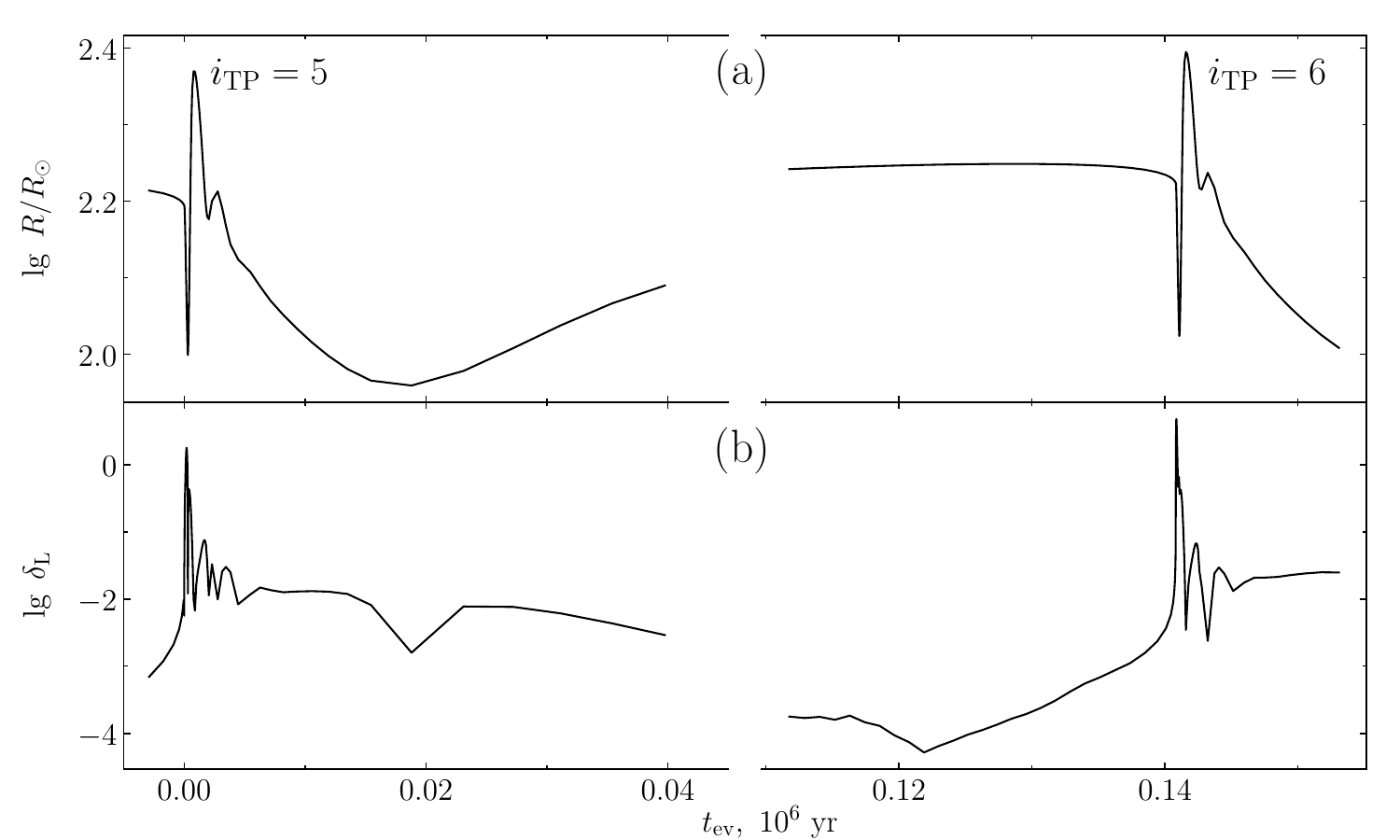}}
\caption{The time dependence of radius of the evolutionary model (a) and deviation from thermal
         equilibrium (b) in the convection zone of the star with initial mass
         $M_\mathrm{ZAMS}=2M_\odot$ during the time intervals comprising the 5--th and
         6--th thermal helium--shell flashes.}
\label{fig1}
\end{figure}

Fig.~\ref{fig1} shows variations of the stellar radius $R$ and parameter $\delta_\mathrm{L}$ for
the evolutionary sequence $M_\mathrm{ZAMS}=2M_\odot$ during the time intervals comprising the fifth
($i_\mathrm{TP}=5$) and the sixth  ($i_\mathrm{TP}=6$) thermal helium--shell flashes.
For the sake of convenience, the evolution time $t_\mathrm{ev}$ is set to zero at the peak helium
shell luminosity
$L_{3\alpha}$ for $i_\mathrm{TP}=5$ when the stellar mass is $M=1.97M_\odot$.
The deviation from thermal equilibrium was calculated for the stellar envelope with the inner boundary
at the Lagrangean coordinate $M_r=0.526M_\odot$.
Within the time interval shown in Fig.~\ref{fig1} the radius and temperature of the envelope inner
boundary varies within $0.073R_\odot\le r_0\le 4.66R_\odot$ and
$6.1\times 10^5~\textrm{K}\le T_0\le 2.3\times 10^7~\textrm{K}$, respectively.
As seen in Fig.~\ref{fig1}, the stellar envelope is in thermal imbalance
(i.e. $\delta_\mathrm{L} > 0.01$) during $\approx 1.4\times 10^4$ yr or
$\approx 10\%$ of the interflash period.
Thus, the models in thermal equilibrium allow us to reproduce the period--luminosity dependence
with a sufficiently good accuracy for $\approx 90\%$ of Mira lifetime.

\section*{hydrodynamic models of miras}

Hydrodynamic models of pulsating AGB stars are computed via solution of the Cauchy problem
for equations of radiation hydrodynamics describing the spherically--symmetric motions of
the stellar envelope.
The convective luminosity and the flux of the mean specific turbulent energy are
calculated with the help of the transport equations for the time--dependent turbulent convection
(Kuhfu\ss\ 1986).
The general system of the equations of hydrodynamics and parameters of the convection theory are
discussed in our earlier paper (Fadeyev 2013).

The initial conditions for the Cauchy problem were computed for selected evolutionary models with
deviation from thermal equilibrium in the outer convection zone $\delta_\mathrm{L} < 10^{-2}$.
The inner boundary of the hydrodynamic model was set in layers with the temperature
$10^6\,\textrm{K}\lesssim T\lesssim 10^7\,\textrm{K}$ and radius $r_0\lesssim 10^{-3}R$,
where $R$ is the radius of the outer boundary of the evolutionary model.
The Lagrangean grid of the hydrodynamic model is computed for mass intervals $\Delta M_r$
increasing geometrically inward.
Significant difficulties in the hydrodynamic calculations of Mira models appear because of
sharply increasing pressure and temperature gradients in the layers with temperature
$T\gtrsim 10^6$~K.
To avoid the large approximation errors at the bottom of the convection zone we
used the Lagrangean grid with 100 innermost mass intervals decreasing geometrically inwards.
The boundary between the regions with increasing and decreasing mass intervals locates in layers
with temperature $T\sim 10^5$~K.

Evolutionary computations of AGB star models were done with the number of mass zones
$3\times 10^3\lesssim N_\mathrm{MESA}\lesssim 5\times 10^3$ whereas the number of mass
zones in the hydrodynamic models of stellar envelopes was in the range $600\le N\le 900$.
The larger number of mass zones was used in pulsation calculations of late AGB stars
because of their high amplitude oscillations.
Initial values of the variables at the grid nodes were calculated using the nonlinear
interpolation of the evolutionary model data with respect to Lagrangean coordinate $M_r$.

\begin{figure}
\centerline{\includegraphics[width=0.9\textwidth]{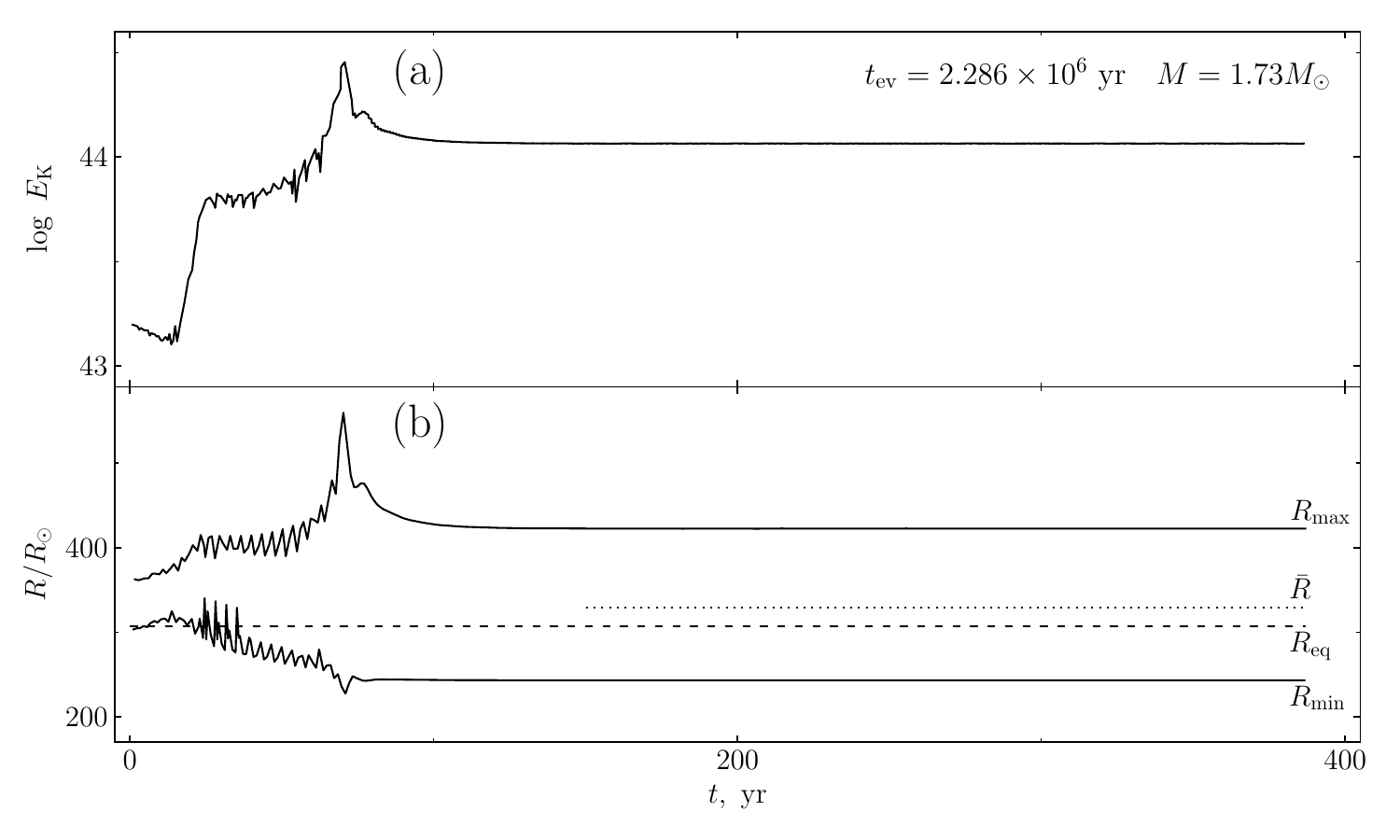}}
\caption{Variations of maximum values of the kinetic energy $E_{\mathrm{K},\max}$ (a)
         as well as maximum $R_{\max}$ and minimum $R_{\min}$ values of the outer boundary radius (b)
         of the hydrodynamic model computed for the evolutionary sequence $M_\mathrm{ZAMS}=2M_\odot$.
         The star age $t_\mathrm{ev}$ yr is set to zero at the first helium--shell flash.
         The stellar mass is $M=1.73M_\odot$.
         The dashed and dotted lines show the radius of the equilibrium evolutionary
         model $R_\mathrm{eq}$ and the mean outer boundary radius of the hydrodynamic model
         $\bar{R}$.}
\label{fig2}
\end{figure}

\begin{figure}
\centerline{\includegraphics[width=0.9\textwidth]{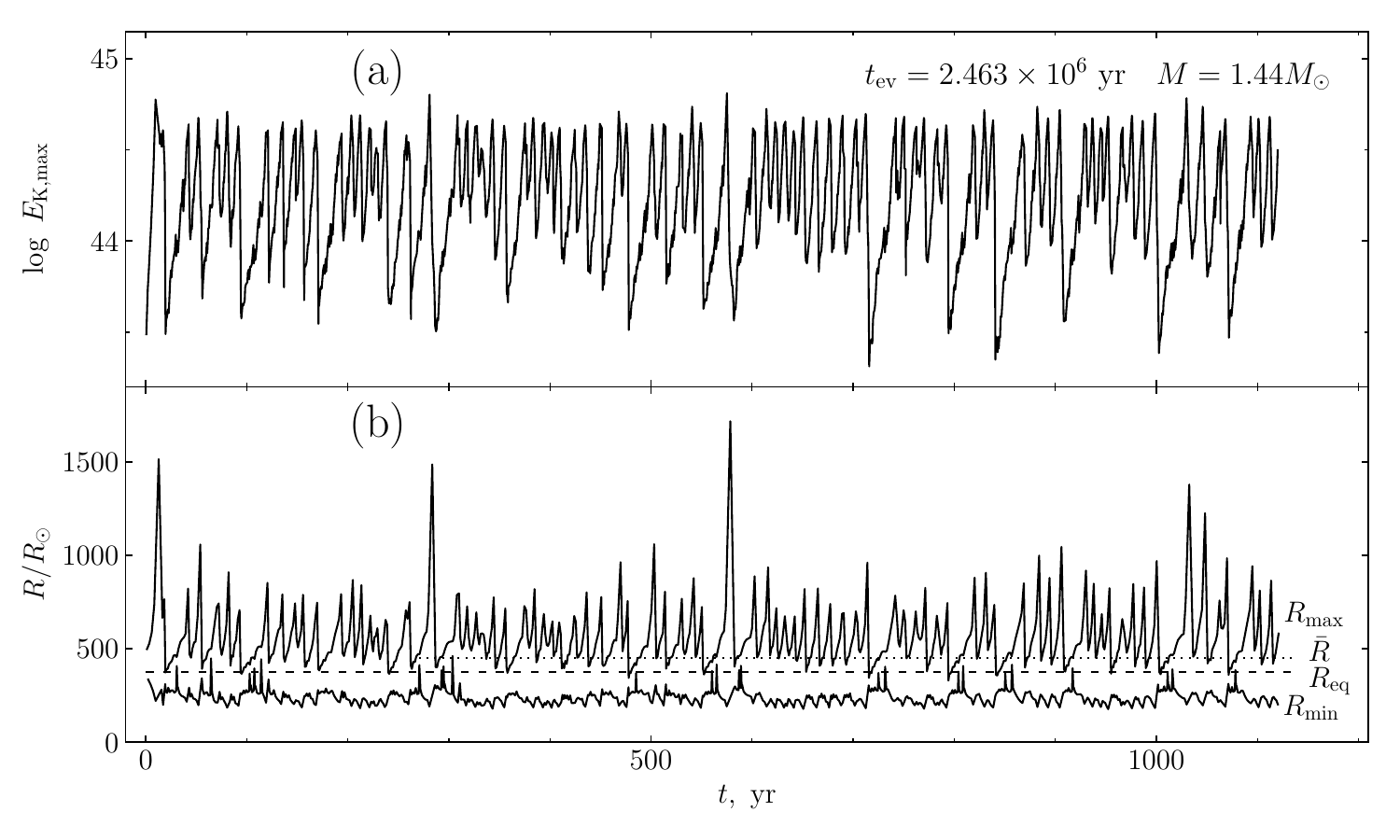}}
\caption{Same as Fig.~\ref{fig2} but for the hydrodynamic model of the AGB star with age
         $t_\mathrm{ev}=2.463\times 10^6$ yr and the mass $M=1.44M_\odot$.}
\label{fig3}
\end{figure}

The solution of the Cauchy problem describes the self--excited stellar oscillations where
the role of small initial perturbations is played by interpolation errors accompanying the
calculation of initial conditions.
The plots illustrating the amplitude growth with subsequent limiting amplitude attainment
are shown in Figs.~\ref{fig2} and \ref{fig3}
for two hydrodynamic models of the evolutionary sequence $M_\mathrm{ZAMS}=2M_\odot$.
For the sake of graphical convenience the results of hydrodynamic computations are illustrated by
the plots of the maximum values of the kinetic energy of pulsation motions $E_{\mathrm{K},\max}$
as well as by the maximum and minimum values of the outer boundary radius $R_{\max}$ and $R_{\min}$.

Both models shown in Figs.~\ref{fig2} and \ref{fig3} are characterized by high amplitude growth
rates whereas the limiting amplitude attainment is not only due to saturation of the
$\kappa$--mechanism responsible for excitation of oscillations, but also due to abrupt
dissipation of the kinetic energy by shock waves appearing in the outer layers of the
hydrodynamic model.
Different amplitude growth rates
($\eta = \Pi d\ln E_{\mathrm{K},\max}/dt = 0.032$ for the model in Fig.~\ref{fig2} and
$\eta = 0.42$ for the model in Fig.~\ref{fig3}) are due to extension of the hydrogen
ionizing zone and increasing pulsation instability as a star evolves along the AGB.

The most striking feature of the plots in Figs.~\ref{fig2} and \ref{fig3} is variations
of the kinetic energy and radius of the outer boundary after the cessation of the
instability growth.
In the first case (see Fig.~\ref{fig2}) the solution converges to fairly regular oscillations
with nearly constant amplitude whereas in the second case (see Fig.~\ref{fig3})
the regular oscillations are not attained because of the large amplitude of the outer layers
even despite significantly longer integration with respect to time.

The mean stellar radius after the cessation of the instability growth $\bar{R}$ was
evaluated with the help of the discrete Fourier transform and as is seen in Figs.~\ref{fig2}
and \ref{fig3}, the estimates of $\bar{R}$ noticeably exceed the equilibrium stellar radius
$R_\mathrm{eq}$.
The difference between $\bar{R}$ and $R_\mathrm{eq}$ is due to the large amplitude oscillations
and periodic shock waves that lead to distension of the outer stellar envelope (Willson 2000).
In the model shown in Fig.~\ref{fig2} the ratio of the mean radius to the equilibrium radius
is $\bar{R}/R_\mathrm{eq} = 1.07$ whereas the mean amplitude of the radial displacement at
the outer boundary is $\Delta R/\bar{R}=0.52$.
In the model shown in Fig.~\ref{fig3} these quantities are
$\bar{R}/R_\mathrm{eq} = 1.21$ и $\Delta R/\bar{R}=0.65$.

\begin{figure}
\centerline{\includegraphics[width=0.9\textwidth]{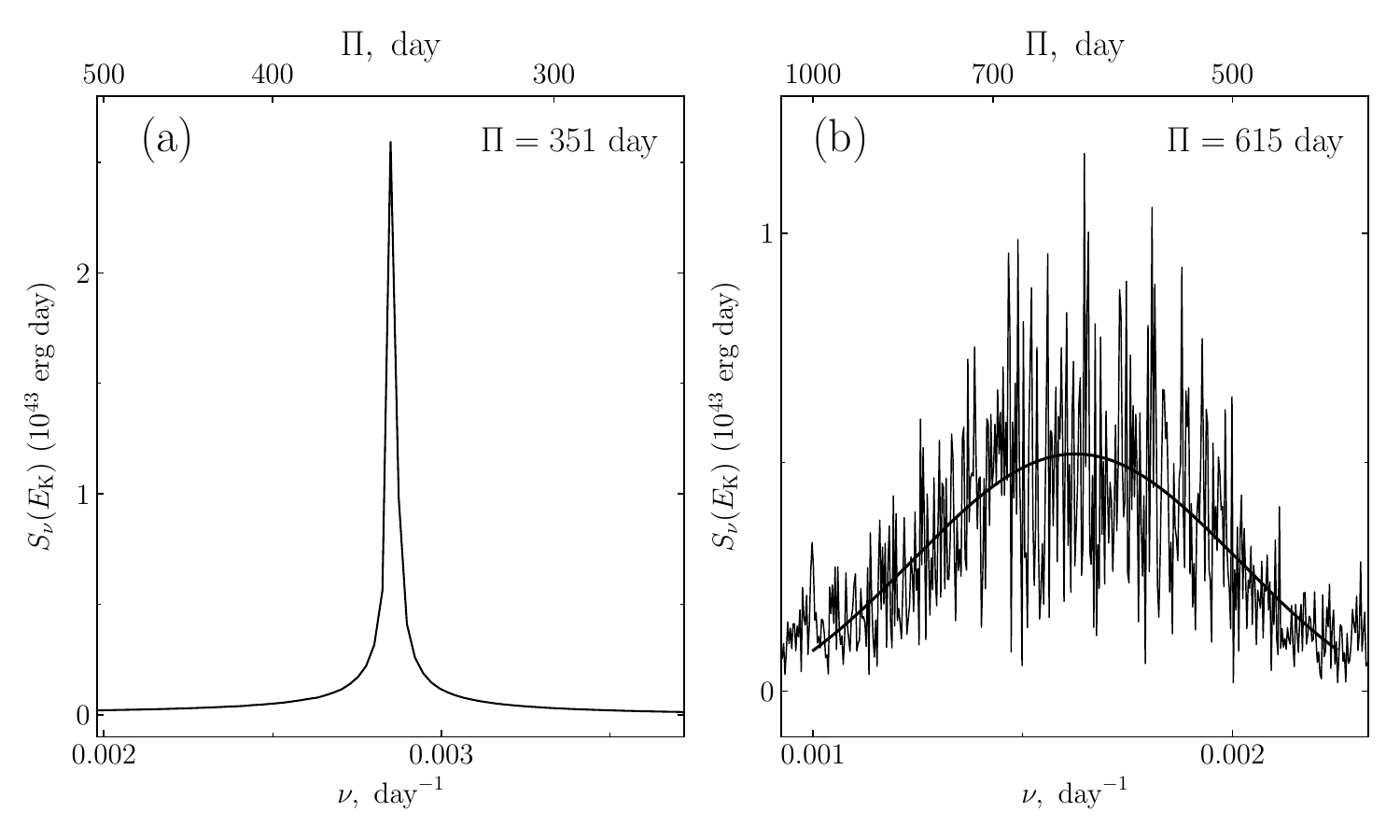}}
\caption{The spectral density of the kinetic energy of pulsation motions $S(E_\mathrm{K})$
         for hydrodynamic models shown in Figs.~\ref{fig2} and \ref{fig3}.
         The least square fit of the spectral density by the gaussian function is shown by
         the thick solid line.}
\label{fig4}
\end{figure}

The mean pulsation period was calculated with the help of the discrete Fourier transform of the
kinetic energy of pulsation motions after attainment of limiting amplitude.
The method of period evaluation is illustrated in Fig.~\ref{fig4} where the plots of the spectral
density of the kinetic energy $S_\nu(E_\mathrm{K})$ in the vicinity of the main oscillation frequency
are shown for two hydrodynamic models with different behavior of limiting amplitude oscillations.
The plot in Fig.~\ref{fig4}a corresponds to the hydrodynamic model shown in Fig.~\ref{fig2}
and the period is easily determined from the maximum of $S_\nu(E_\mathrm{K})$.
The plot of the spectral density in Fig.~\ref{fig4}b corresponds to irregular oscillations
and the mean period is evaluated from the maximum of the Gaussian function fitting the
spectrum $S_\nu(E_\mathrm{K})$ using the least--squares method.

To obtain the reliable estimates of the mean pulsation period $\Pi$ we carried out the
calculations of stellar pulsations on the intervals comprising more than $10^3$ periods.
However one should bear in mind that the cessation of amplitude growth does not necessarily
imply that the model reaches the limiting amplitude oscillations because cessation of the
amplitude growth is usually followed by the model relaxation with slowly changing mean
values of $\bar{R}$ and luminosity $\bar{L}$.
Results of preliminary calculations allow us to conclude that a criterion of limiting
amplitude oscillations is $\bar{L} = L_0$, where $L_0$ is the total luminosity at the inner boundary of
the hydrodynamic model.
In the present study the calculations of hydrodynamic models were ended as soon as
the time interval of the discrete Fourier transform reaches $t/\Pi\approx 10^3$ and
the difference between $L_0$ and the mean luminosity $\bar{L}$ does not exceed 0.1\%.

\section*{period--radius and period--luminosity relations}

\begin{figure}
\centerline{\includegraphics[width=0.9\textwidth]{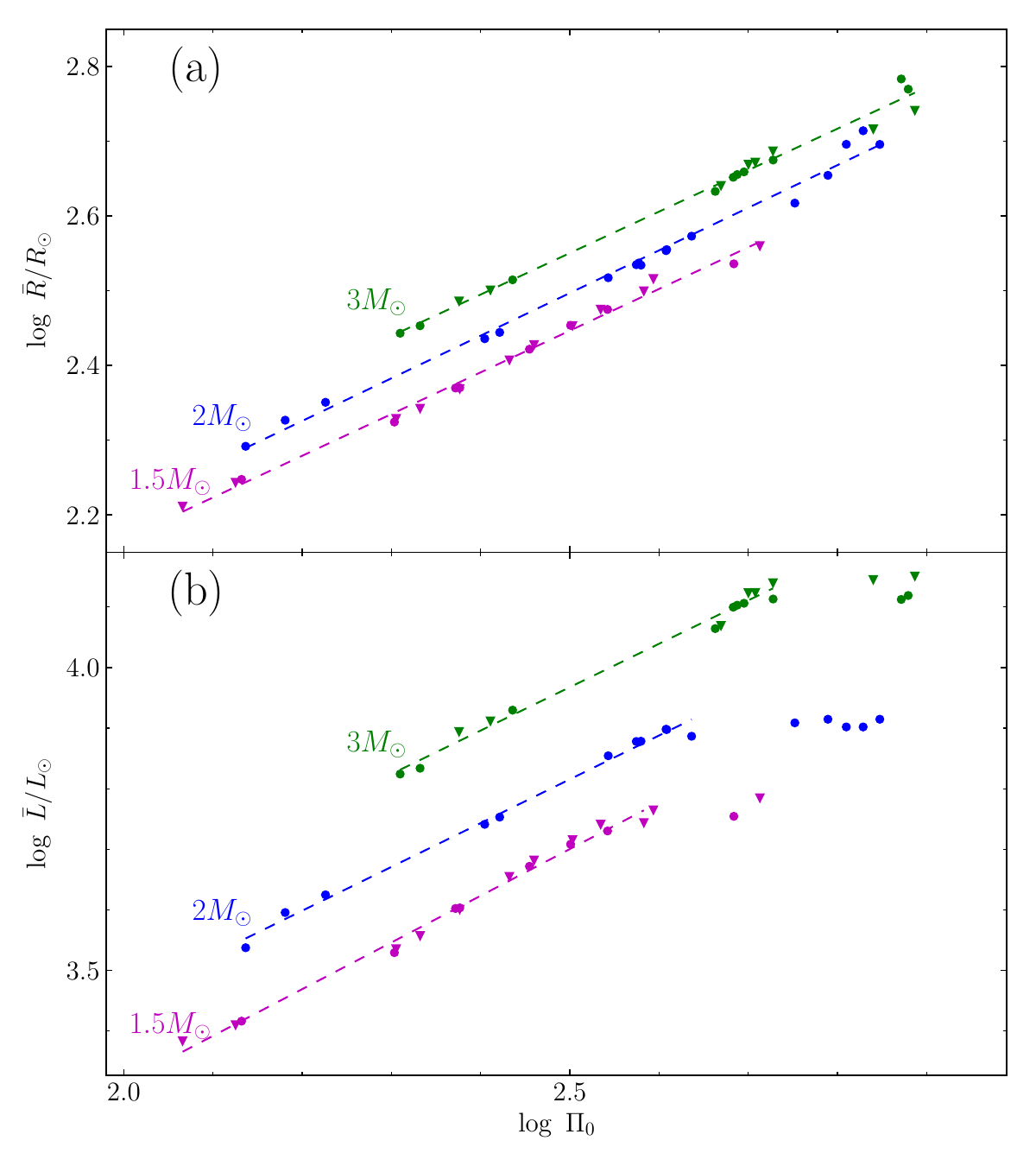}}
\caption{The period--radius (a) and period--luminosity (b) relations obtained from computations
         of hydrodynamic models of fundamental mode pulsators.
         The circles and triangles correspond to the models of evolutionary sequences
         computed with the mass loss rate parameters $\eta_\mathrm{B}=0.05$ and
         $\eta_\mathrm{B}=0.03$, respectively.}
\label{fig5}
\end{figure}

\begin{figure}
\centerline{\includegraphics[width=0.9\textwidth]{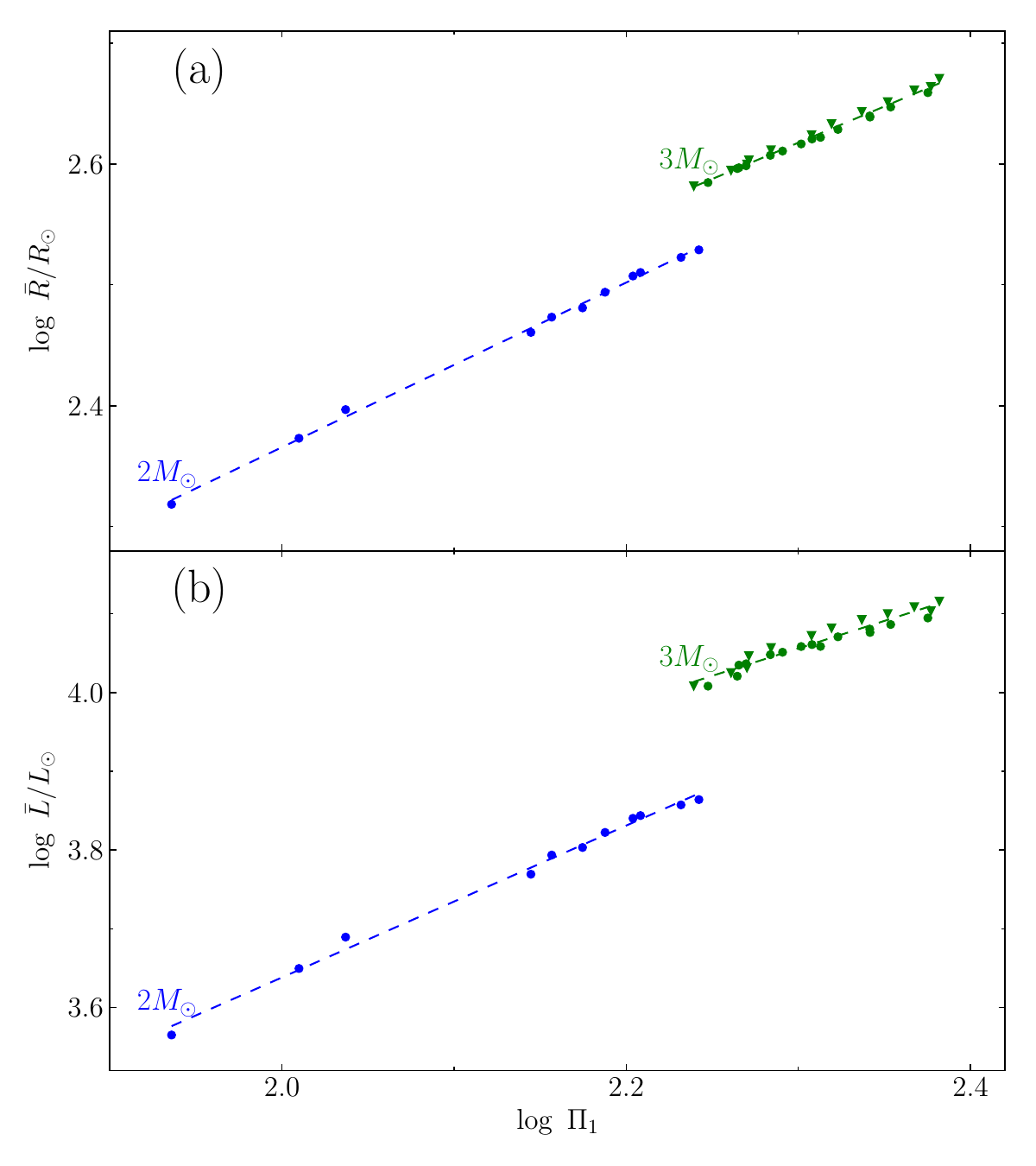}}
\caption{Same as in Fig.~\ref{fig5} but for hydrodynamic models of the first overtone pulsators.}
\label{fig6}
\end{figure}

In this work we computed 150 hydrodynamic models of Mira variables.
Nearly 30 models were found to be stable against radial oscillations since they showed
the decaying amplitude.
The period--radius $\Pi-\bar{R}$ and period--luminosity $\Pi-\bar{L}$ dependencies
determined with the help of hydrodynamic models are shown in Figs.~\ref{fig5} and \ref{fig6}
for pulsations in the fundamental mode and first overtone, respectively.

As mentioned above, the evolutionary sequences $M_\mathrm{ZAMS}=1.5M_\odot$ and $3M_\odot$ were
computed with the mass loss rate parameters $\eta_\mathrm{B}=0.03$ and $\eta_\mathrm{B}=0.05$.
In Figs.~\ref{fig5} and \ref{fig6} the corresponding hydrodynamic models are shown by the
triangles and circles.
It is seen that the models of each evolutionary sequence are concentrated along the
continuous line and are almost independent of $\eta_\mathrm{B}$.

The dashed lines in Figs.~\ref{fig5} and \ref{fig6} represent the linear least--squares fits
\begin{gather}
\label{p-r}
\lg\bar{R}/R_\odot = a_0 + a_1 \lg\Pi_k ,
\\
\label{p-l}
\lg\bar{L}/L_\odot = b_0 + b_1 \lg\Pi_k ,
\end{gather}
where $k=0$ and $k=1$ correspond to the fundamental mode and first overtone pulsations,
respectively.
The coefficients $a_0$, $a_1$, $b_0$, $b_1$ in (\ref{p-r}) and (\ref{p-l})
as well as the ranges of corresponding period intervals ($\Pi_a, \Pi_b$)
are listed in Table~\ref{tabl1}.

As seen in Fig.~\ref{fig5}, the scatter of points becomes larger with increasing $\Pi$ and
indicates an increase in errors of the pulsation period estimates for $\Pi > 500$ d.
Moreover, the models with longest periods show the systematic deviations in the period--luminosity
diagram to smaller values of $\bar{L}$ though the similar deviations in the period--radius diagram
are not present.

The systematic deviations in Fig.~\ref{fig5}b are due to the strong non--linearity
of long period stellar pulsations.
On the other hand, the systematic deviations in the period--radius diagram are not present
because the increase of non--linearity is accompanied by increase both in period and radius
because they relate as $\Pi\propto\bar{R}^{3/2}$.
Similar transformations in the $\Pi-\bar{L}$ diagram are impossible because the mean stellar
luminosity remains constant so that increase of stellar pulsation period with increasing
pulsation amplitude leads to the shift along the horizontal axis of the diagram.

In the present study the period--radius and period--luminosity relations were determined for
the models of evolutionary sequences $M_\mathrm{ZAMS}=1.5M_\odot$, $2M_\odot$ and $3M_\odot$
pulsating in the fundamental mode and first overtone within the whole period interval.
The only exception is the period--luminosity relation for the fundamental mode pulsators
where the period interval is limited to the linear dependence between $\log\Pi$ and $\log \bar{L}$.

\section*{conclusion}

The theoretical period--radius and period--luminosity relations for Mira models with
solar metallicity were determined on the basis of consistent calculations of stellar evolution
nonlinear stellar pulsations.
This method is applicable for the stellar envelopes under thermal equilibrium and
allows us to consider $\approx 90\%$ of these stars.
Results of computations show that the theoretical period--radius and period--luminosity
relations depend on the initial mass of the star on the main sequence.
For example, the fundamental mode oscillations with period $\Pi=400$ d appear in models
of evolutionary sequences $M_\mathrm{ZAMS}=1.5M_\odot$ and $M_\mathrm{ZAMS}=3M_\odot$
with the luminosity difference $\Delta m_\mathrm{bol}\approx 0.7$ mag.
Dependence of $\Pi-\bar{R}$ and $\Pi-\bar{L}$ on $M_\mathrm{ZAMS}$ is due to the fact that
the total luminosity of AGB stars is a function of the mass of the degenerate carbon core
and monotonically increases with increasing core mass
(Paczynski, 1970; Uus 1970).
Therefore, the dispersion of the observable relations $\Pi-R$ and $\Pi-L$ is not only due to
observational errors but also due to different initial masses of observed stars.
It is also clear that the relations $\Pi-\bar{R}$ and $\Pi-\bar{L}$ computed for the
mass loss rate parameters $\eta_\mathrm{B}=0.03$ and $\eta_\mathrm{B}=0.05$
insignificantly differ from one another.
Indeed, the pulsation period and stellar mass relate as $\Pi\propto M^{-1/2}$ and therefore
decrease in stellar mass during the initial AGB phase does not affect the
pulsation period and plays a perceptible role only during the short final AGB stage.

It is noteworthy to mention that the periods of the first overtone pulsators range in rather
narrow intervals: $86~\textrm{d}\le\Pi\le 123~\textrm{d}$ and $174~\textrm{d}\le\Pi\le 204~\textrm{d}$
for $M_\mathrm{ZAMS}=2M_\odot$ and $M_\mathrm{ZAMS}=3M_\odot$, respectively.
More dense grids of evolutionary sequences and hydrodynamic models will probably help
to identify the pulsation modes of Mira variables.

The main difficulties in hydrodynamic modeling of Mira variables arise in computations of
late AGB star models when the large oscillation amplitude leads to bad convergence of
iterative solution of implicit difference equations.
A method for overcoming this obstacle implies enlargement of the number of Lagrangean zones
in the hydrodynamic model  with simultaneous reduction of the integration time step.
Unfortunately, this approach leads to a significant increase in computational time.
Moreover, the problem becomes more difficult because in order to obtain a reliable estimate
of the period of large amplitude oscillations the equations of hydrodynamics should be solved
on the longer time interval.

More severe difficulties arise in computation of Mira hydrodynamic models with
periods $\Pi\sim 10^3$~d corresponding to the final evolutionary phase of AGB.
In the present study we failed to obtain the stable solution of the equations of
hydrodynamics because even small computational errors lead to the dynamical instability
and expansion of the outer Lagrangean zones with velocity exceeding the local escape
velocity.
The problem is that the envelope of the late--phase AGB star is near the boundary
of dynamical instability because of the widely extended hydrogen ionizing zone where the
adiabatic exponent is below its critical value ($\Gamma_1 = (\partial\ln P/\partial\ln\rho)_S < 4/3$)
whereas the surface gravity becomes too low due to the large stellar radius.
A similar opinion has been earlier expressed by Tuchman et al. (1978; 1979) in their
discussion of non--linear stellar pulsation as the cause of planetary nebula formation.

\newpage
\section*{references}

\begin{enumerate}

\item M. Andriantsaralaza, S. Ramstedt, W.H.T. Vlemmings, and E. De Beck,
      Astron. Astrophys. \textbf{667}, A74 (2022).

\item M. Asplund, N. Grevesse, A.J. Sauval and P. Scott,
      Ann. Rev. Astron. Astrophys. \textbf{47}, 481 (2009).

\item T. Bl\"ocker, Astron. Astrophys. \textbf{297}, 727 (1995).

\item E. B\"ohm--Vitense, Zeitschrift f\"ur Astrophys. \textbf{46}, 108 (1958).

\item J.O. Chibueze, R. Urago, T. Omodaka, Yu. Morikawa, M.Y. Fujimoto, A. Nakagawa,
      T. Nagayama, T. Nagayama, and K. Hirano, Publ. Astron. Soc. Japan \textbf{72}, 59 (2020).

\item R.H. Cyburt, A.M. Amthor, R. Ferguson, Z. Meisel, K. Smith, S. Warren, A. Heger,
      R.D. Hoffman, T. Rauscher, A. Sakharuk, H. Schatz, F.K. Thielemann, and M. Wiescher,
      Astrophys. J. Suppl. Ser. \textbf{189}, 240 (2010).

\item Yu.A. Fadeyev, Astron. Lett. \textbf{39}, 306 (2013).

\item Yu.A. Fadeyev, MNRAS \textbf{514}, 5996 (2022).

\item M.W. Feast, MNRAS \textbf{211}, 51 (1984).

\item M.W. Feast, Observatory \textbf{105}, 85 (1985)

\item M.W. Feast, I.S. Glass, P.A. Whitelock and  R.M. Catchpole, MNRAS \textbf{241}, 375 (1989).

\item I.S. Glass and T. Lloyd Evans, Nature \textbf{291}, 303 (1981).

\item I.S. Glass and M.W. Feast, MNRAS \textbf{198}, 199 (1982а).

\item I.S. Glass and M.W. Feast, MNRAS \textbf{199}, 245 (1982б).

\item M.A.T. Groenewegen and P.A. Whitelock, MNRAS \textbf{281}, 1347 (1996).

\item F. Herwig, Astron. Astrophys. \textbf{360}, 952 (2000).

\item C.D. Huang, A.G. Riess, S.L. Hoffmann, Ch. Klein, J. Bloom, W. Yuan, M. M. Lucas,
      D.O. Jones, P.A. Whitelock, S. Casertano, and R.I. Anderson, Astrophys. J. \textbf{857}, 67 (2018).

\item R. Kuhfu\ss, Astron. Astrophys. \textbf{160}, 116 (1986).

\item J. Mould, J.R. Graham, K. Matthews Keith, G. Neugebauer, and J. Elias,
      Astrophys. J. \textbf{349}, 503 (1990).

\item J. Mould, A. Saha Abhijit, and S. Hughes, Astrophys. J. Suppl. Ser. \textbf{154}, 623 (2004).

\item B. Paczy\'nski, Acta Astron. \textbf{20}, 47 (1970).

\item B. Paxton, R. Smolec, J. Schwab, A. Gautschy, L. Bildsten, M. Cantiello, A. Dotter,
      R. Farmer, J.A. Goldberg, A.S. Jermyn, S.M. Kanbur, P. Marchant, A. Thoul, R.H.D. Townsend, W.M. Wolf,
      M. Zhang, and F.X. Timmes, Astrophys. J. Suppl. Ser. \textbf{243}, 10 (2019).

\item M. Pignatari, F. Herwig, R. Hirschi, M. Bennett, G. Rockefeller, C. Fryer,
      F.X. Timmes, C. Ritter, A. Heger, S. Jones, U. Battino, A. Dotter, R. Trappitsch, S. Diehl,
      U. Frischknecht, A. Hungerford, G. Magkotsios, C. Travaglio, and P. Young,
      Astrophys. J. Suppl. Ser. \textbf{225}, 24 (2016).

\item D. Reimers, \textit{Problems in stellar atmospheres and envelopes} (Ed. B. Baschek, W.H. Kegel,
      G. Traving, New York: Springer-Verlag, 1975), p. 229.

\item Y. Sun, B. Zhang, M.J. Reid, Sh. Xu, Sh. Wen, J. Zhang, and X. Zheng,
      Astrophys. J \textbf{931}, 74 (2022).

\item Y. Tuchman, N. Sack, and Z. Barkat, Astrophys. J \textbf{219}, 183 (1978).

\item Y. Tuchman, N. Sack, and Z. Barkat, Astrophys. J \textbf{234}, 217 (1979).

\item R. Urago, R. Yamaguchi, T. Omodaka, T. Nagayama, J.O. Chibueze, M.Y. Fujimoto,
      T. Nagayama, A. Nakagawa, Yu. Ueno, M. Kawabata, T. Nakaoka, K. Takagi, M. Yamanaka, K. Kawabata,
      Publ. Astron. Soc. Japan \textbf{72}, 57 (2020).

\item U. Uus, Nauchn. Inform. Astron. Sovet AN SSSR \textbf{17}, 25 (1970).

\item P.A. Whitelock and M.W. Feast, MNRAS \textbf{319}, 759 (2000).

\item P.A. Whitelock, F. Marang Freddy and M.W. Feast, MNRAS \textbf{319}, 728 (2000).

\item P.A. Whitelock, J.W. Menzies, M.W. Feast, F. Nsengiyumva, and N. Matsunaga,
      MNRAS \textbf{428}, 2216 (2013).

\item L.A. Willson, Annual Rev. Astron. Astrophys \textbf{38}, 573 (2000).

\item W. Yuan, M.M. Lucas, A. Javadi, Zh. Lin, and J.Z. Huang, Astron. J. \textbf{156}, 112 (2018).

\item H. Zhang and J.L. Sanders, MNRAS \textbf{521}, 1462 (2023).
\end{enumerate}

\newpage
\begin{table}
\caption{Parameters of the period--radius (\ref{p-r}) and period--luminosity (\ref{p-l}) relations.}
\label{tabl1}
\begin{center}

\begin{tabular}{cccccccccc}
\hline
$M_\mathrm{ZAMS}/M_\odot$  & $k$ & $a_0$ & $a_1$  & $\Pi_a$ & $\Pi_b$ & $b_0$ & $b_1$ & $\Pi_a$ & $\Pi_b$ \\
\hline
1.5 & 0 &  1.051 &  0.558 &  116 &  245 &  1.962 &  0.689 &  116 &  245 \\
2.0 & 0 &  1.110 &  0.554 &  137 &  309 &  2.009 &  0.723 &  137 &  244 \\
    & 1 &  1.003 &  0.681 &   86 &  123 &  1.709 &  0.964 &   86 &  123 \\
3.0 & 0 &  1.183 &  0.547 &  204 &  397 &  2.196 &  0.709 &  204 &  330 \\
    & 1 &  1.237 &  0.600 &  174 &  204 &  2.451 &  0.698 &  174 &  204 \\
\hline
\end{tabular}

\end{center}
\end{table}
\clearpage

\newpage
\clearpage

\newpage
\clearpage

\newpage
\clearpage

\newpage
\clearpage

\newpage
\clearpage

\newpage
\clearpage

\end{document}